\documentclass[12pt,preprint]{aastex}
\usepackage{amsmath}
\usepackage{amssymb}
\begin{document}

\title{On The Orbital Evolution of Jupiter Mass Protoplanet Embedded in \\
       A Self-Gravity Disk}

\author{Hui Zhang \altaffilmark{1,2},Chi Yuan\altaffilmark{1},
D.N.C. Lin\altaffilmark{3,4}and David C.C. Yen\altaffilmark{1,5}}

\altaffiltext{1}{Institute of astronomy and astrophysics, Academia
Sinica ,Taipei} \altaffiltext{2}{Department of Astronomy,NanJing
University, Nanjing} \altaffiltext{3}{UCO/Lick Observatory,
University of California, Santa Cruz} \altaffiltext{4}{Kavli
Institute of Astronomy and Astrophysics, Peking University,Peking}
\altaffiltext{5}{Department of Mathematics, Fu Jen Catholic
University, Taipei}

\begin{abstract}

We performed a series of hydro-dynamic simulations to investigate
the orbital migration of a Jovian planet embedded in a
proto-stellar disk. In order to take into account of the effect of
the disk's self gravity, we developed and adopted an
\textbf{Antares} code which is based on a 2-D Godunov scheme to
obtain the exact Reimann solution for isothermal or polytropic
gas, with non-reflecting boundary conditions. Our simulations
indicate that in the study of the runaway (type III) migration, it
is important to carry out a fully self consistent treatment of the
gravitational interaction between the disk and the embedded
planet. Through a series of convergence tests, we show that
adequate numerical resolution, especially within the planet's
Roche lobe, critically determines the outcome of the simulations.
We consider a variety of initial conditions and show that
isolated, non eccentric protoplanet planets do not undergo type
III migration.  We attribute the difference between our and
previous simulations to the contribution of a self consistent
representation of the disk's self gravity. Nevertheless, type III
migration cannot be completely suppressed and its onset requires
finite amplitude perturbations such as that induced by
planet-planet interaction.  We determine the radial extent of type
III migration as a function of the disk's self gravity.

\end{abstract}


\section{Introduction}
More than 200 planets have been discovered around nearby solar-type
stars.  Their kinematic properties are characterized by diversities
in their mass, period, eccentricity, and physical radius.  An
important dynamical process which may have led to these properties
is protoplanets' migration due to their tidal interaction with their
nascent disks \citep{lin93,pap06}. The progenitor cores of gas giant
planets undergo type I migration due to a torque imbalance between
different regions of the disks where they are embedded
\citep{gol80,war84}. After they have acquired sufficient mass to
open a gap near their orbit, gas giant planets' orbital evolution is
locked to that of the disk gas through a type-II migration
\citep{lin86}.  But under some circumstances, the gap may be
partially cleared and the disk gas which leak through this region
can induce the gas giants to undergo runaway (type III) migration
(Masset \& Papaloizou 2003, hereafter MP).

All of these process can relocate protoplanets far from their birth
place. The rapid time scale for type-I migration \citep{war97,tan02}
poses a challenge to the formation of gas giant planets
\citep{tho06,ida07}. But several potential retardation mechanisms
have been proposed. They include variation in the surface density
and temperature gradient \citep{mas06b}, intrinsic turbulence in the
disk \citep{lau04,nel04}, and non-linear radiative and hydrodynamic
feedbacks \citep{mas06a}. If these cores are formed in the turbulent
free dead zone \citep{gam96}, self-induced unstable flow
\citep{kol04,li05,dev07} would reduce the efficiency of type I
migration by an order of magnitude (Dobbs-Dixon {\it et al.} 2007,
in preparation).

Type-II migration has been invoked as a mechanism for the formation
of close-in gas giants \citep{lin96}.  Both 1D \citep{lin86} and 2D
\citep{ang06} simulations have shown that before disk gas mass
decays to the value comparable to the planet mass, the planet
migrates with (unperturbed) disk accretion on a viscous diffusion
time scale and when disk gas mass is comparable to the planet mass,
only a fraction of the total (viscous plus advective) angular
momentum flux transported by the disk gas (which is assumed to be
independent of the disk radius) is utilized by the planet in its
orbital evolution \citep{iva99}. The inclusion of type-II migration
in the planet formation models has yield a mass-period distribution
which is similar to that observed (Ida \& Lin 2004a,2004b, 2005,
2007).

Type III migration is driven by a strong corotation torque very near
the planet (Ida {\it et al.} 2000, MP). The time scale for this
process is much shorter than both type I and type II migration.  If
it commonly occurs, type III migration would greatly erase any
signature in the dynamical structure of planetary systems from the
disk initial surface density distribution of their nascent disks.
Due to its dramatic effects, type III migration has been extensively
studied over the past few years \citep{ogi06}. This process can only
be maintained if there is an uninterrupted flow across the planet's
orbit so that saturation of the corotation resonances can be
avoided. Masset and Papaloizou(2003) show that the radial motion of
a rapidly migrating planet can indeed be self sustained, {\it i.e.}
its motion leads to the fresh supply to the corotation region which
provides a torque to induce further migration. Ida {\it et al.} 2000
demonstrated a similar effect for planetary migration induced by
residual planetesimals. They suggest that the critical condition for
the onset of this process is that the mass of the residual
planetesimals contained within the feeding zone (with a half width
up to a few times that of the Roche radius) must exceed that of the
planet.  The results of the hydrodynamic simulations (MP) also show
that a planet would undergoes runaway migration in disk regions with
$\delta m \geq M_{planet}$, where $\delta m$ is the mass of the disk
gas in the planet's co-orbital region.  For a Jupiter-mass, this
requirement implies a mass ratio between the entire disk and the
central star to be $\mu\equiv\frac{M_{d}}{M_{\star}} \ga 0.02$.

In such disks, the effect of the disk self gravity is important,
especially in the determination of the torque applied to the gas
in the planet's co-orbital region by that in other regions of the
disk. But in the previous numerical simulations, the effect of the
disk's self gravity has been neglected.  In this paper, we develop
a new numerical scheme which takes the effect of disk self gravity
into account.  With this scheme, our main objectives are to
examine the conditions under which type III migration is launched
and sustained. In \S2, we provide a description of our
computational method and model parameters.

The simulation of MP showed that type III migration is spontaneous
excited in relatively massive disks.  We consider two limiting
initial conditions.  In the first set of simulations, we consider
the emergence of an isolated planet on a circular orbit.  In the
core accretion scenario, the growth of gas giants occur on time
scales much longer than the synodic time scale in most nearby disk
regions such that the stream lines can adjust adiabatically.  In
the second set of simulations, we introduce a set of ``dynamically
quiescent'' initial conditions by gradually increasing the
planet's mass over many orbits and the initial angular velocity of
planet is not exact Keplerian(The centrifugal force will balances
the gravity both from center star and the whole gas disk) so that
the disk gas can adjust to its tidal potential of the planet and
attain a dynamical equilibrium. During this transition stage, the
semi major axis of the planet is artificially held fixed. This
``quiet-start'' prescription is introduced to minimize the impulse
felt by the gas at the onset of the simulation.  With this initial
condition, we carried out several series of simulations to
determine the dependence of the disk flow and planet's migration
on the numerical resolution and the degree of the disk self
gravity.  In \S3, we present the results of these simulations.

The ``quiet-start'' models provide a test on whether type-III
migration may be spontaneously launched under the optimum conditions
as indicated by the numerical simulation by MP.  They also provided
a powerful analytic argument to suggest that it can be sustained
once type-III migration is initiated. Ida {\it et al.} (2000) found
that this process may need an ``initial push'' due to some large
perturbations. In the second series of simulations, we consider this
possibility by introducing an initial jolt as a trigger for type-III
migration.  One potential mechanism for this impulsive perturbation
is close encounter between two proto-planet\citep{zho05}. In \S4, we
present the simulated results of the ``initial-push'' models.
Finally, we summary our results and discuss their implications in
\S5.

\section{Physical and Numerical Model}

\subsection{Physical model}
Following conventional procedures, we simulate the dynamical
response of a gas disk around a star which is located at the origin
of the coordinates.  We constructed a 2D numerical hydrodynamic
scheme to solve the continuity and momentum equations, neglecting
the effect of any explicit viscosity.  We place a protoplanet which
is initially embedded in the disk with a circular orbit around the
central star. In order to avoid some well-known problems (see below)
at the inner boundary (close to the central star), we solve the
governing equations in the Cartesian coordinate.

\begin{figure}
\epsscale{1} \plotone{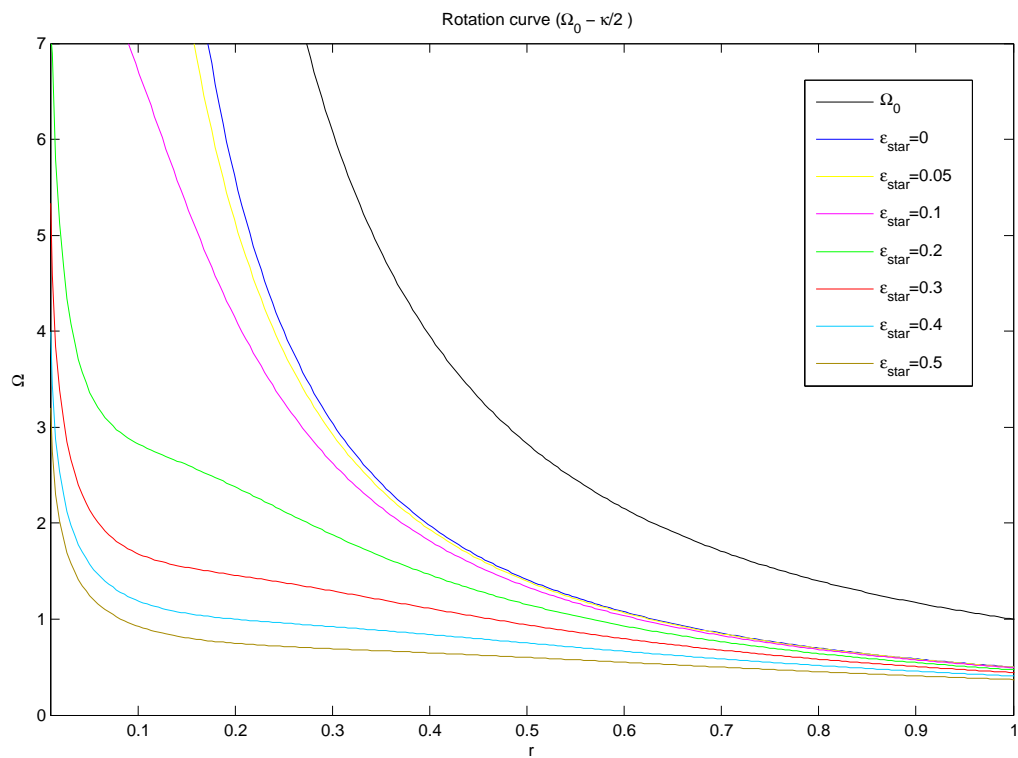} \caption{Rotation curves of disk. We
test the rotation curves when different $\epsilon_{star}$ is
adopted. The curve on the top is $\Omega_0$ when $\epsilon_{star}=0$
and curves below it are $\Omega_0-\frac{\kappa}{2}$ when
$\epsilon_{star}=0,0.05,0.1,0.2,0.3,0.4,0.5$. In our simulations
$\epsilon_{star}=0.1$, and in most part of the disk the rotation
curve is very close to the one when $\epsilon_{star}=0$. The
difference is it reaches finite value instead of going infinite when
$r\rightarrow 0$. \label{figure 1}}
\end{figure}

The vertically averaged continuity equation for the disk gas is given
by
\begin{equation}
\frac{\partial \sigma}{\partial t} + \frac{\partial(\sigma
u_{x})}{\partial x} + \frac{\partial(\sigma u_{y})}{\partial y} =
0
\end{equation}
The equations of motion in the Cartesian coordinates are
\begin{equation}
\frac{\partial(\sigma u_{x})}{\partial t}+\frac{\partial(\sigma
u_{x}^{2})}{\partial x}+\frac{\partial(\sigma u_{x}
u_{y})}{\partial y} = -\frac{\partial P}{\partial x}-\sigma
\frac{\partial \Phi}{\partial x}
\end{equation}

\begin{equation}
\frac{\partial(\sigma u_{y})}{\partial t}+\frac{\partial(\sigma
u_{x} u_{y})}{\partial x}+\frac{\partial(\sigma
u_{y}^{2})}{\partial y} = -\frac{\partial P}{\partial y}-\sigma
\frac{\partial \Phi}{\partial y}
\end{equation}
where P is pressure and $\Phi$ is the gravity potential of the
star-planet-disk system, which includes the softened potential of
central star ($\Phi_s$), softened potential of the planet
($\Phi_p$), potential of the disk itself ($\Phi_d$) and indirect
potential ($\Phi_i$) due to the acceleration of origin by planet and
disk. The softened potential of central star is given by
$\Phi_{s}=-\frac{GM_\odot}{\sqrt{x^2+y^2+\epsilon_{star}^2}}$ and
the softened potential of planet is
$\Phi_{p}=-\frac{GM_p}{\sqrt{(x-x_p)^2+(y-y_p)^2+\epsilon_{p}^2}}$
where $\epsilon_{star}$ and $\epsilon_{p}$ are the soften length to
central star and planet respectively.  In all the models presented
here, we adopt $\epsilon_{p}$ is half of the planet's Roche radius.
\textbf{Fig}.\ref{figure 1} shows the rotation curves of the disk
when we adopt different $\epsilon_{star}$. During our simulations it
is 0.1 in units where the initial semi major axis of the planet is
unity (in some other simulations we reduce it to 0.05 and find
little difference).

\subsection{Numerical Method}
The \textbf{Antares} code we have developed is adopted in the
calculations. It is a 2-D Godunov code based on the exact Riemann
solution for isothermal or polytropic gas, featured with
non-reflecting boundary conditions. The details of this code has
described elsewhere\citep{yua05}.

Full self-gravity of the disk is calculated by FFT.  We assume the
disk gas has an isothermal equation of state and we don't add any
explicit viscosity in the simulation.  There is however some
numerical viscosity associated with our computational scheme.
Extensive tests indicate that the magnitude of artificial
viscosity is equivalent to $\alpha<10^{-4}$. Shock is treated with
standard von Neuman prescription.  For the orbit of planet we
adopt RK78 to integrate it.

\subsection{Computational mesh configuration and domain}
In many models, the planet's orbit undergoes extensive decays.  A
natural system to solve the governing equations is the polar
coordinates.  But, the axial symmetric of the disk flow is broken by
the presence of the planet.  The inner boundary conditions can only be
an approximate function of the disk radius along well inside the orbit
of the planet.  Although the Fargo prescription provides a resolution
for this technical challenge over some regions of the disk, it is
nonetheless difficult to achieve extend the computational domain to
very small $R_{inner}$ with Polar coordinates because the
computational time $T_{com} \sim R_{inner}^{-3/2}$.

\begin{figure}
\epsscale{1} \plotone{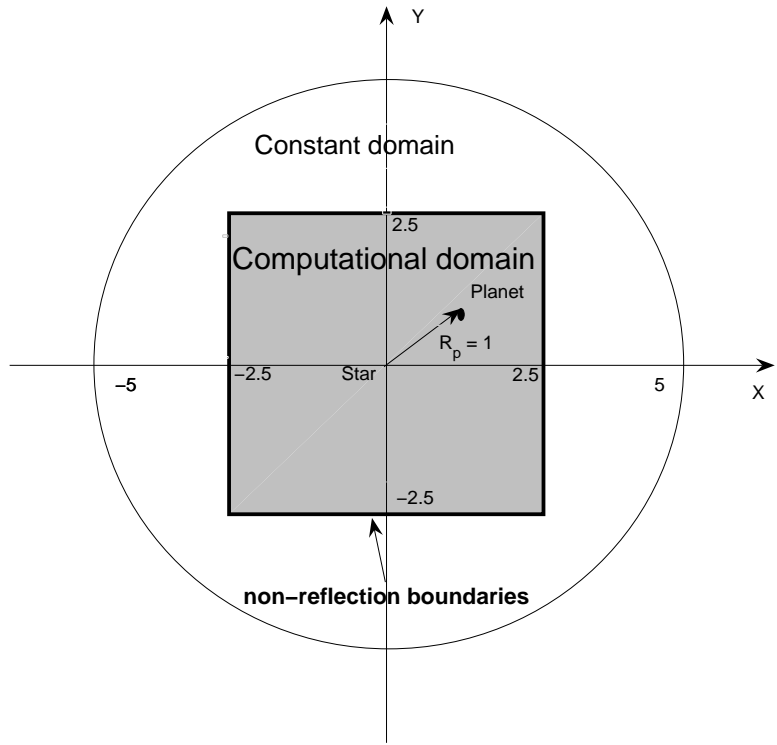} \caption{Computational domain.
Computational domain is from -2.5 to 2.5 in x direction and from
-2.5 to 2.5 in y direction(gray square). Surrounding it are four
non-reflection boundaries. Area outside the square is assume to stay
constant. \label{figure 2}}
\end{figure}

A Cartesian coordinate system introduces some advantages over the
polar coordinates. It is easy to achieve high resolution without the
bottle-neck in the azimuthal direction.  It is also straight forward
to calculate self-gravitating effect with FFT. During some of our
simulations we adopt a $512\times512$ grid, while in some high
resolution cases it increases to $1024\times1024$. To carry out the
calculation, however, a softening length $\epsilon_{star}$ is
assigned to the central star, where $\epsilon{star}$ is one order
smaller than the length unit. Computational domain is from -2.5 to
2.5 in both x,y direction. Primary star locates at the origin where
x=0 and y=0. To avoid symmetric problem of the four corners we
extend the disk to R=5 and assume the area outside the computational
square($x \times y=5\times5$) will stay constant(See
\textbf{Fig}.\ref{figure 2}) during the evolution, while the gravity
is taken account all over the area within $R\leq5$.

\subsection{Computational Units}
For numerical convenience we set gravitational constant $G=1$, solar
mass $M_{\odot}=1$ and the radius of planet's initial orbit $R_0=1$,
where $R_0=5.2AU$. The unit of time is $1/2\pi$ of the planet's
initial orbit period $P_0$. Most of our simulations are carried out
over a period of $1000P_0$. Since the unit of density $\sigma_0$ drops
out of the equations of motion, we can normalize it to any specified
density.

\subsection{Initial Conditions}
At the beginning of evolution, the disk surface density is set to
be uniform. We adopt 4 different initial surface density:
$\Sigma_0=0.6\times10^{-3}$,$0.9\times10^{-3}$,$1.2\times10^{-3}$,
$1.5\times10^{-3}$.  The mass of thinnest disk is about
$0.012M_{\odot}$ which is about the minimum mass of solar nebula,
while the thickest one is about $0.03M_{\odot}$. The angular
velocity of the gas $v_\theta=r \Omega_g$ is slightly different
from the Keplerian velocity since the flow is in a centrifugal
balance with both the softened gravity of the star and
self-gravity of the disk (when self-gravitating effect is
included) such that $v_{\theta} =\Omega_g
r=\sqrt{\frac{rGM_\odot}{r^2+ \epsilon^2} +rf_{sg}}$.  In disks
with an isothermal equation of state and a homogenous surface
density distribution, the pressure gradient effect does not
contribute to the initial azimuthal speed.  The initial radial
velocity of gas is set to be 0. These initial disk conditions do
not take into account the gravitational perturbation by the
planet.

The initial azimuthal velocity of the planet is also set to balance
the gravity of the central star and that of the disk. The initial
location of planet is at (x=1,y=0).  In the simplest models, planet's
initial mass is $10^{-3}$ ($1M_J$) and is fixed during the
evolution. This approach introduces a gravitational impulse which can
strongly perturb the stream line, especially near the planet's orbit.

For ``quiet-start'' models, we adjust the initial velocity to set
up a dynamical equilibrium in which the planet's orbit is circular
and the stream lines are closed.  To do so, we adopt an negligible
initial mass for the planet ($3\times10^{-7}$ or equivalently
0.1$M_\oplus$). We specify the planet's growth rate to be $3\%$
during every orbit period until it grows to 1$M_J$ within the
first 250 fixed circular orbit periods. With this prescription the
planet gains mass through adiabatic growth and the disk has enough
time to make smooth response. Before the planet is launched, we
further adjusted the azimuthal speed of the planet so that it
regains a circular orbit despite the presence of the gas.

\subsection{Boundary conditions}
We adopt non-reflection boundary condition, that means we do wave
decomposition at each boundary and set all the waves that propagate
inward to computational domain to be zero. So the wave can only
propagate outward and we had assume the area is constant outside
boundary, that makes the wave absorbed at the boundary. While the
boundary isn't closed to mass flow, gas may flows through the
boundary freely according to the equation of motion. The area
outside the boundary is assume to be uniform and maintain the
initial condition without evolution. The details of this boundary
condition had been described elsewhere\citep{god96}.

\section{Simulations of flow with a quiet starts}

A series of hydrodynamic simulations had been performed(see
\textbf{Table}~\ref{table 1}). At first, we consider a series of
simplest models(S1-S4).  In these models, the numerical resolution is
relatively low ($512\times512$), the prescription for self-gravitating
and quiet-start effects are NOT included.  We specify the planet's
mass to one Jupiter mass and test four different surface density of
disk.

\begin{figure}
\epsscale{1} \plotone{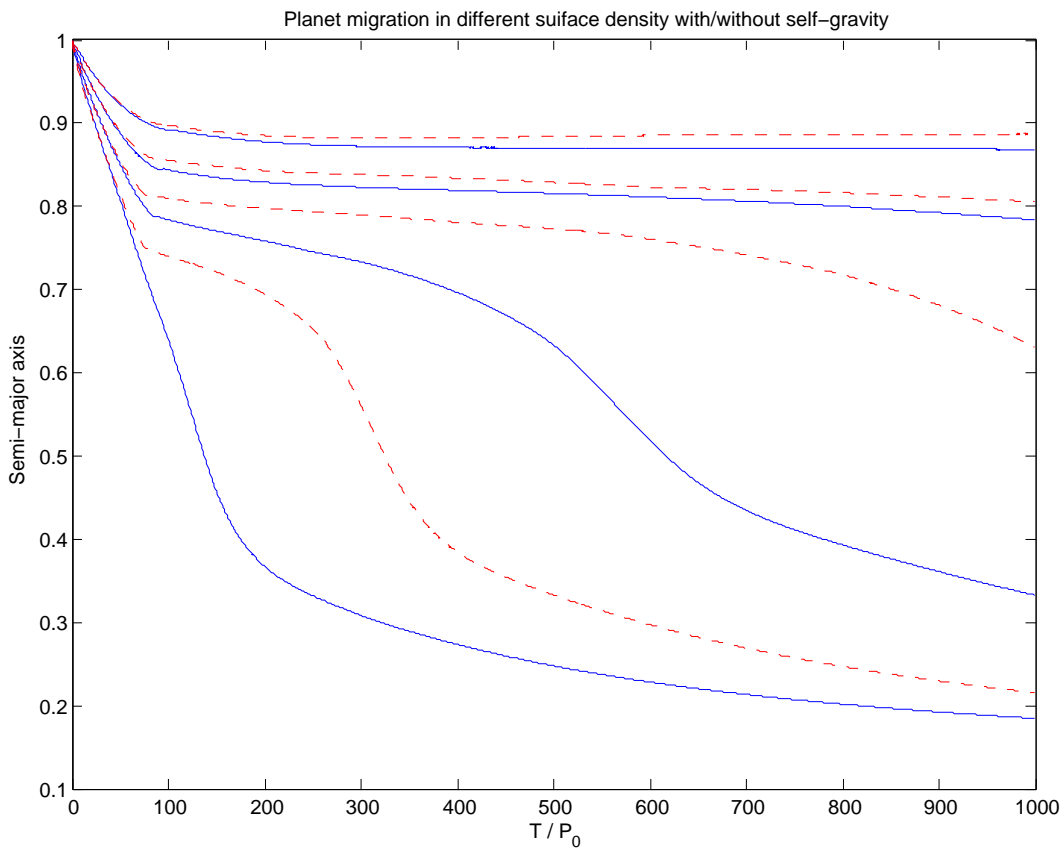} \caption{Migration in different disk
surface density. From top to bottom,
$\sigma=0.6\times10^{-3}$,$0.9\times10^{-3}$,$1.2\times10^{-3}$,$1.5\times10^{-3}$.MMSN
mass corresponds to $0.6\times10^{-3}$ in our units. The blue solid
lines are the cases without self-gravity while the red dash lines
denote the cases with self-gravity. The critical density to trigger
runaway migration becomes higher in a self-gravitating
disk.\label{figure 3}}
\end{figure}

From the lowest to the highest surface density $\Sigma=0.6\times
10^{-3}$,$0.9\times 10^{-3}$,$1.2\times 10^{-3}$ and $1.5\times
10^{-3}$ (in units of solar mass divides by the square of the planet's
initial radius). The results show that the migration rate is
proportional to the surface density of disk.  In the limit that the
disk's surface density is higher than $1.2\times10^{-3}$, a very rapid
migration occurs(See \textbf{Fig}~\ref{figure 3}).

This result is in agreement with that obtained by MP. For the critical
model, the disk-to-primary mass ratio ($\pi \Sigma R_d^2 / M_\ast$) is
a little above 0.01 and planet-to-primary mass ratio is 0.001.
According to the criterion specified by MP, this set of parameter is
at the boundary of the 'runaway domain', so the migration curve shows
a critical property. When the surface density becomes higher, the
model parameters are totally in the 'runaway area' and the migration
is much faster. MP had suggested that this rapid migration is due to
the planet's co-rotation torque being consistently replenished by
the disk gas which flow through the planet's orbit.  We present detailed
analysis for the high resolution models to support with conjecture.

\subsection{Resolution}
The above simulation show that the torque density is most intense near
the orbit of the protoplanet, especially within its Roche radius.
With the adopted mass ratio, the planet's Roche radius is about 0.069
in our dimensionless unit. For the relatively low resolution
($512\times512$) models, the width of each grid is $\delta x=\delta y=
0.01$.  In these models, there are only 7 grids within planet's Roche
radius.  Torques associated with large $m$ resonances can not be well
resolved in this limited resolution and the under-resolved torques may
lead to some unreasonable effect.

In principle, the nearly symmetric flow pattern within the Roche
radius is expected to lead to large cancellation of the net torque
applied by the circum planetary disk on the planet at its center.
But the direct consequence of inadequate numerical resolution is
that the gas accumulated in the Roche lobe will generate large
artificial fluctuations in the magnitude of the torque applied to
the planet which cannot be easily cancelled.  In Figure \ref{figure
4}, we show that the ratio ($\Gamma$) of the tidal torque (on the
planet) by the gas within planet's Roche lobe to the tidal torque
(on the planet) by the gas within the entire disk. The rapid
oscillations of this ratio is clearly shown.  This inadequate
resolution of the flow pattern introduces an inconsistency in which
the planet and the gas flow within it's Roche lobe are dragged along
by each other.

\begin{figure}
\epsscale{1} \plotone{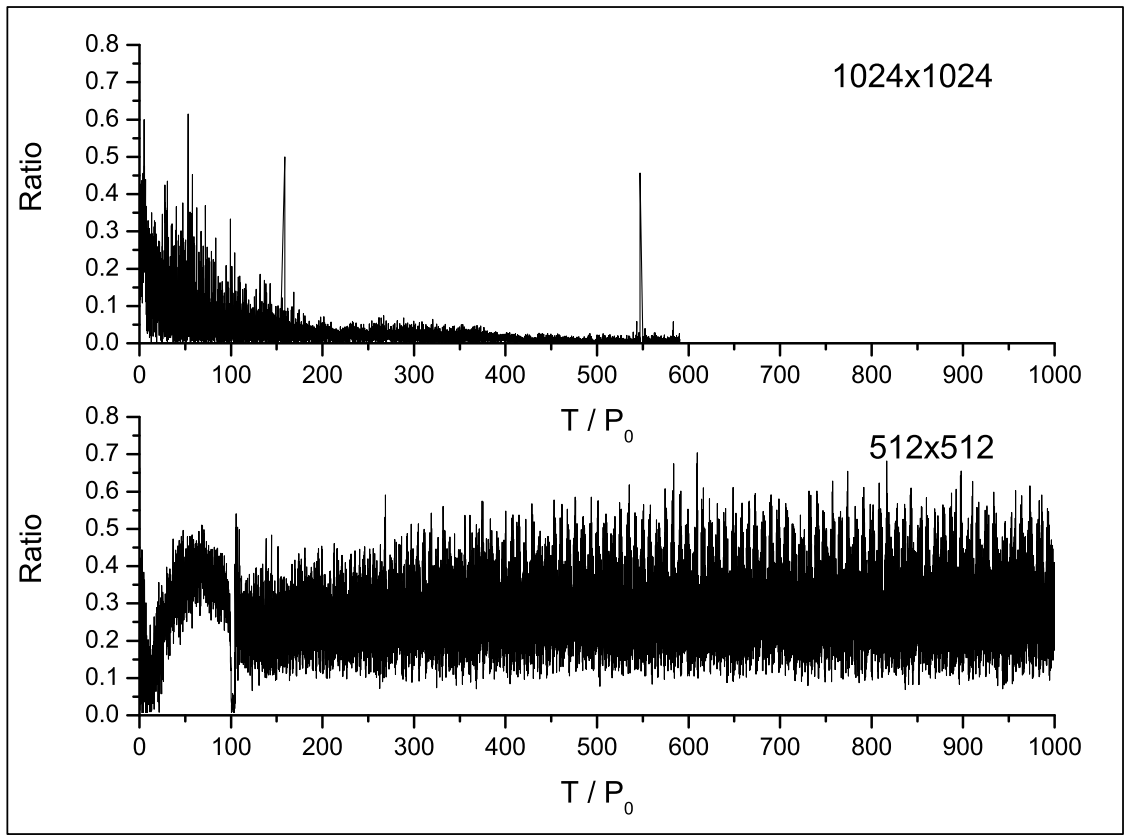} \caption{Ratio of torque(to planet)
within planet's Roche lobe to the torque(to planet) within the whole
disk. The top one's resolution is $1024\times1024$ and the bottom is
of $512\times512$. In the low resolution case the torque within
planet's Roche lobe is almost half of the whole torque during the
simulation. And the rapid, large oscillations correspond to the
unbalanced torque which is the consequence of mass flowing across
unresolved vicinity of the planet. In high resolution case, the
problem is not very serious. \label{figure 4}}
\end{figure}

The issue with resolving the flow within the Roche lobe is clearly
illustrated by the sensitive dependence of the migration rate on the
softening length for the planet's potential (Nelson \& Benz
2003a,2003b; Cresswell \& Nelson 2006). In the low-mass limit where
the Roche radius is smaller, the lack of resolution introduces even
more severe problems for both type-I and type-III migration because
they are strongly determined by the flow close to the planet.
Inadequate resolution is less serious for type-II migration in which
case a gap is clearly formed and the gas in planet's Roche lobe is
depleted. Nevertheless, inadequate resolution may also lead to
artificial diffusion of gas into the gap.  Figure \ref{figure 4}
shows large fluctuation in the magnitude of $\Gamma$ after the gap
has formed (at $\sim 100P_{0}$) in the low resolution models.  There
is also gas flow across planet's Roche lobe where torque imbalance
is amplified by the coarsely resolved mesh.  Leakage of fresh gas
into the planet's co-orbital region can also sustain a gradient in
the potential vorticity and suppression of corotation saturation
\citep{mas06a} which may reduce the efficiency of type II migration
from disk gas accretion\citep{cri07}. The corotation torque scales
with the gradient of the potential vorticity(Goldreich \& Tremaine
1979;Ward 1991,1992):
\begin{equation}
\Gamma_{C}\propto \Sigma\frac{d\log{\frac{\Sigma}{B}}}{d\log{r}},
\end{equation}
For a sufficiently smooth,monotonic transition of surface density
from $\Sigma_i$ to $\Sigma_o$ the vorticity logarithmic gradient is
therefore\citep{mas06a}:
\begin{equation}
\frac{d\log{\frac{\Sigma}{B}}}{d\log{r}}=\frac{r}{\lambda}\log{\frac{\Sigma_o}{\Sigma_i}}(a+b\frac{H^2}{\lambda^2}),
\end{equation}
where $a$ and $b$ are constants of numerical functions of $r$ of
unity that depend on the shape of the surface density profile. $H$
is the scale hight of the disk and $\lambda\ll r$ is the length
scale of the density transition, which are both constant too.
\subsection{Convergence tests}
In order to highlight the problems introduced by the inadequate
numerical resolution, we carry out several high resolution
($1024\times1024$) simulations.  The upper panel of Figure
\ref{figure 4} shows that the fluctuation in the magnitude of
$\Gamma$ declined greatly after the gap has formed (at $\sim
100P_{0}$) in the high resolution models.  A combination in the
reduction of artificial numerical viscosity and adequately
resolved torque greatly reduces the artificial torque imbalance on
the planet due to the gas within the Roche lobe (also see D'Angelo
{\it et al.} 2005).  Consequently, the planet's migration is also
significantly reduced.

\begin{figure}
\epsscale{1} \plotone{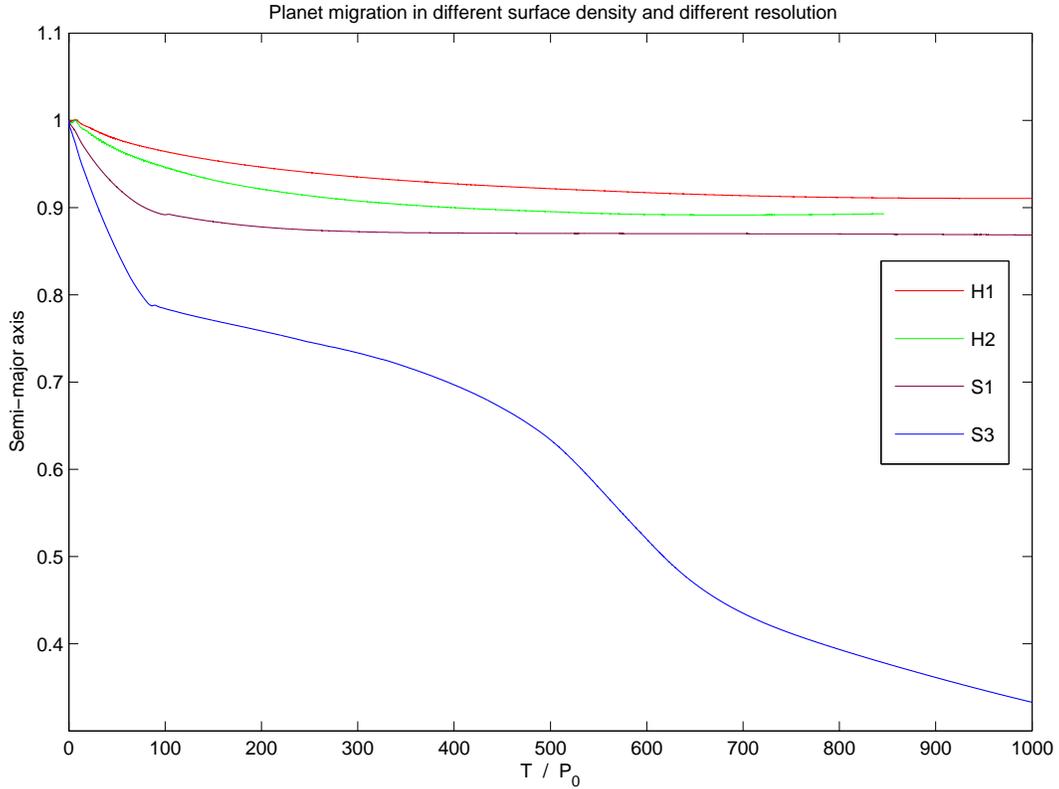} \caption{Migration curves.Red and
green lines show the simulation in the resolution of
$1024\times1024$(H1 and H2) while the purple and blue lines are in
the resolution of $512\times512$(S1 and S3). The red and purple
lines denote normal surface density $\sigma=0.6\times10^{-3}$, the
green and blue lines denote doubled surface density
$\sigma=1.2\times10^{-3}$. 'Runaway migration' occurs in low
resolution case but doesn't show up in high resolution case although
the surface density is the same. \label{figure 5}}
\end{figure}

Figure \ref{figure 5} shows the planet's orbital evolution for
simulations with different resolutions. In two separate sets of
initial surface densities, other than a modest initial radial decay
during the epoch of gap formation, type III (runaway) migration is
essentially eliminated.  The dichotomy between these models (S3 and
H2) is particularly dramatic for the models with sufficiently high
disk mass that the type III migration is spontaneously launched in the
low-resolution simulations.

\begin{figure}
\epsscale{1} \plotone{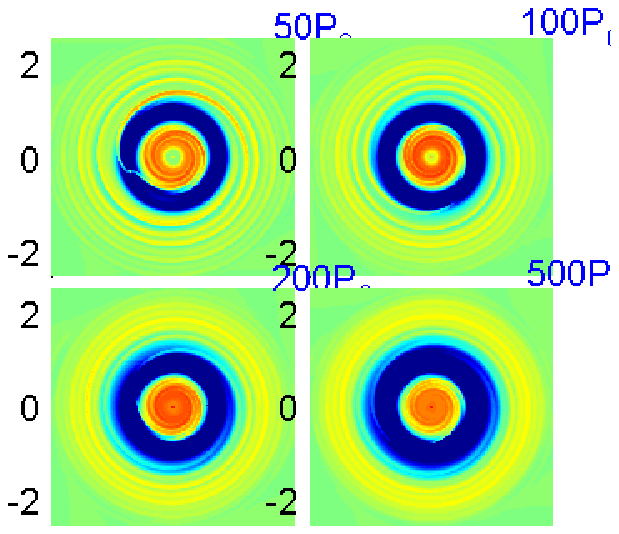} \caption{Density evolution of S3. The
resolution is $512\times512$ and without self-gravity. From the top
left to bottom right, evolution time is $50P_0$, $100P_0$, $200P_0$,
$500P_0$. \label{figure 6}}
\end{figure}

\begin{figure}
\epsscale{1} \plotone{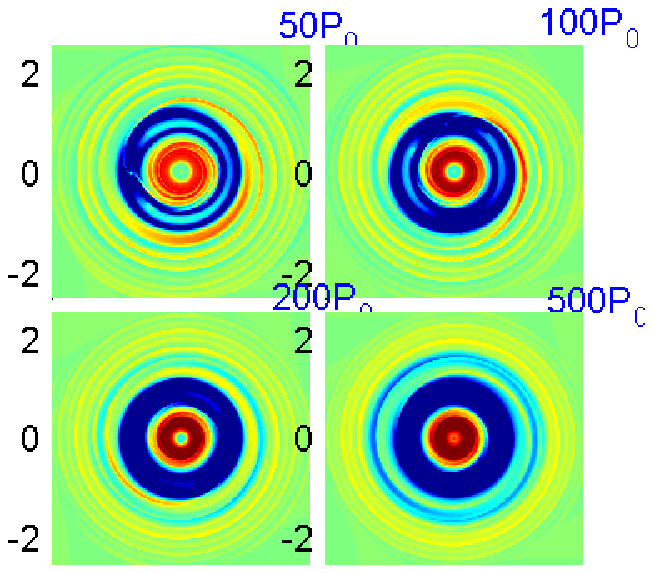} \caption{Density evolution of H2. The
resolution is $1024\times1024$ and without self-gravity. From the
top left to bottom right, evolution time is $50P_0$, $100P_0$,
$200P_0$, $ 500P_0$. \label{figure 7}}
\end{figure}

Another useful diagnostic is the surface density distribution.  In
Figures \ref{figure 6} and \ref{figure 7} we show the density
evolution of models S3 and H2. Figure \ref{figure 7} clearly shows
the presistent presence of gas in the 'horse shoe' region and an
un-axisymmetric structure at the edge of the gap. The high-
resolution simulation in Figure \ref{figure 7} shows a sharper disk
edge and more clearly defined wave pattern than the lower-resolution
simulation. The high resolution simulations require several months
of CPU time.  Although, our simulations are carried out with greater
resolution than most existing calculations, we are not yet able to
achieve higher resolution and test numerical convergence at this
stage.

\subsection{Self-gravity}
The results in the previous subsection illustrate the contribution
to the torque on the planet by the gas within the Roche lobe.
Although this amount of material in this region is small compared
with the mass of the planet, the gravity between it and the planet
is strong due to their proximity.  In fact, a fraction of this gas
actually resides in a disk around and is gravitationally attached
to the planet.  As we have indicated above that the net torque on
the planet by the gas in the proto-planetary disk is expected to
be mostly cancelled and its mass should share the torque from the
rest of the disk with the planet.

In most of the published numerical simulations, the effect of the
disk's self gravity is not included. This approximation introduces an
inconsistent gravitational field felt by the planet and by the gas
which shares its orbit.  In this section, we consider the effect of
self-gravity in models $SG1 \sim SG4$ and $QG1 \sim QG4$.  Due to
limited computational resources, these models are simulated with low
resolutions (512x512).  Although inadequate resolution continues to
plague the proper determination of the tidal torque, we use these
models to demonstrate that a self-consistent treatment of the disk
self gravity couples the flow within the planet's Roche lobe to it and
reduces the rate of type-III migration.

Results in Figure \ref{figure 3} show that the migration slows down
slightly (a few percent) in a low-$\Sigma$ models (SG1 and SG2) when
self-gravitating effect is included (similar results are obtained by
Nelson \& Benz 2003a,b). However in the high-$\Sigma$ models SG3 and
SG4, the difference brought by the self-consistent treatment of the
disk's self gravity is much more pronounced. At $10^3 P_0$, the
migration in case SG3 slows down by almost $50\%$ relatively to that
in model S3 where the effect of the disk's self gravity is
neglected. And more importantly, the 'runaway migration' doesn't
occurs when sufficiently high values of $\Sigma$ which did lead to
'runaway migration' in a non-self-gravitating disk.

We note that, in all models with identical $\Sigma$ distribution,
the planet have the same orbital decay rate regardless whether the
effect of self gravity is included.  This similarity is probably
due to the slightly artificial impulse initial conditions adopted
here. In these series of simulations, gas in the disk is forced to
respond to the planet's gravity for the first time at the onset of
the computation. This initial impulse leads to large potential
vorticity gradient which ensures a strong contribution from the
corotation resonances.  Under some circumstance, the planet
migrates inwards by a sufficiently large increment, a fresh supply
of disk gas with new values of potential vorticity is brought to
the co-orbital region of the planet such that the further
migration is promoted.  For most disks, however, the replenishment
of fresh disk material is inadequate to self sustain the run away
migration.

The difference between these two series of models become more
pronounced after the gap formation.  In this limit, replenishment
of the co-orbital region is quenched.  As gas librate on
horse-shoe orbits, any initial potential vorticity gradient is
erased such that the contribution from the corotation resonance
become saturated and weakened.

\begin{figure}
\epsscale{1} \plotone{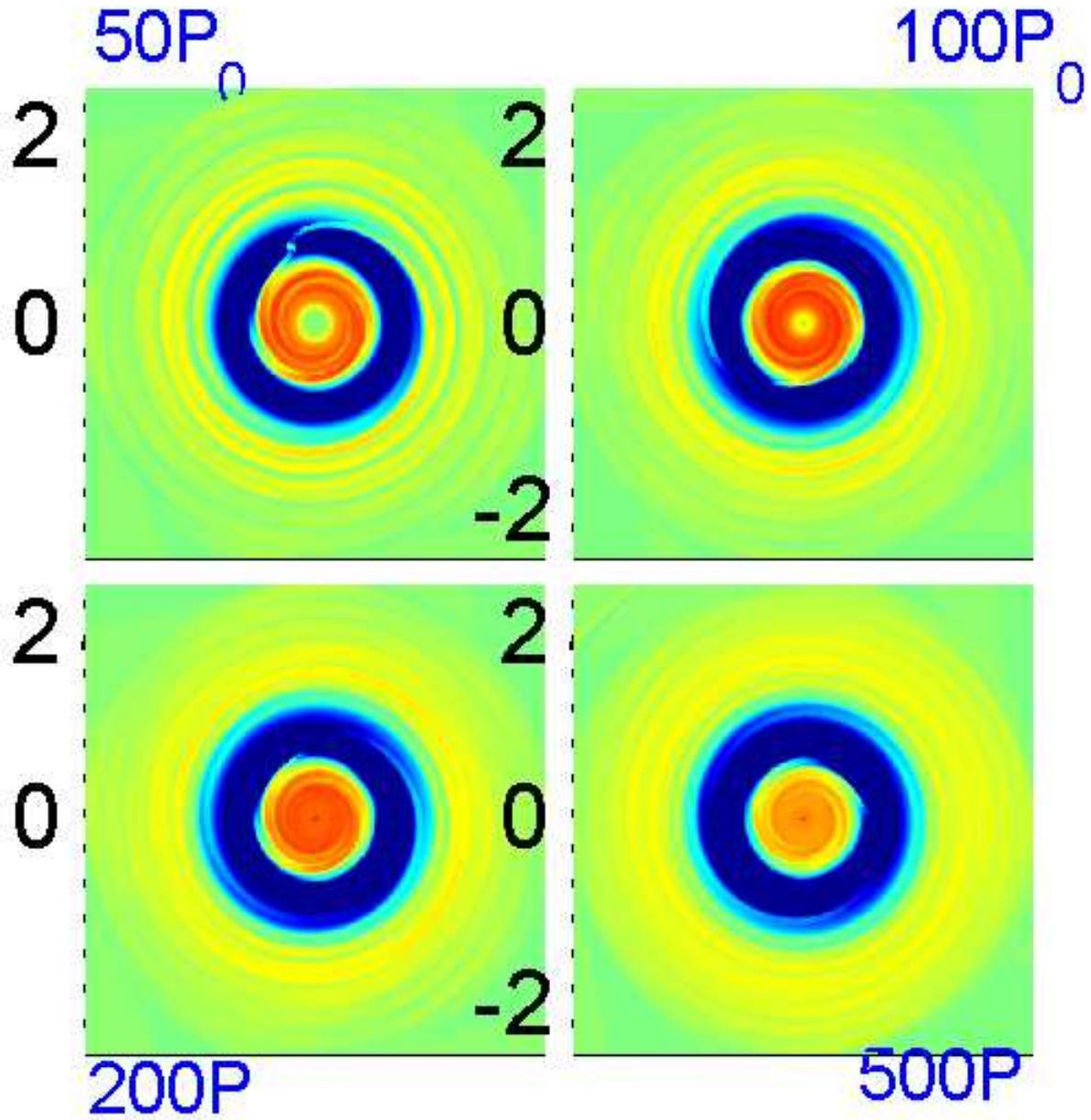} \caption{Density evolution of SG3. The
resolution is $512\times512$ and self-gravitating effect is
included. From the top left to bottom right, evolution time is
$50P_0$, $100P_0$, $200P_0$, $500P_0$. \label{figure 8}}
\end{figure}

In Figure \ref{figure 8}, we plot the $\Sigma$ distribution for the
self gravitating model SG3.  In this model, the clearing of the gap
strongly enhances the effect of self gravity near the boundary of
the gap.  Although the Q-value of the disk near the planet's orbit
is initially $Q\gg1$, but the clearing and accumulation of gas
beyond the gap leads to a local $Q\sim2$ near the outer edge of the
gap.  With such a low Q-value at a relatively sharp disk edge,
un-axisymmetric gravitational\citep{pap89} and
shearing\citep{bal01,li05,dev07} may be excited. These instabilities
can significant reduce the migration speed (Koller {\it et al.}
2003, Dobbs-Dixon {\it et al.} in preparation).

Finally, the treatment of self gravity closely ties together the
planet and the gas within its Roche lobe. With a self consistent
treatment of the gas self gravity, the interaction between the
planet and a significant fraction of the gas within its Roche lobe
becomes a binding rather than dispersive force.  The self gravity
of the gas beyond the gap region can only start to dominant the
torque after the flow has established an equilibrium pattern such
that the gas in the co-orbital region migrate together with the
planet as integral parts.

\section{Isolated protoplanet versus perturbed system}

The results in Figure \ref{figure 3} suggest a transition in the
protoplanet-disk tidal interaction from being dominated by corotation
resonances to their saturation as gas in the co-orbital region is
being cleared out.  We consider two limiting possibilities: a
``quiet'' and an ``impulsive'' start.

\subsection{Quiet start}

According to the core-accretion scenario, the most favorable
location for the first generation of gas giants to form is near the
snow line \citep{ida04a}.  Since the growth time scales for their
progenitor cores is sensitively dependent on their disk environment
and their gas accretion is a runaway process, the first gas giants
are likely to form in isolation over many dynamical time scales
\citep{pol96}. In principle, the disk can adjust adiabatically to
the perturbation due to the emergence of the gas giants.  This
expectation provides the rational for a set of simulations with a
``quiet start'' (for a description of the quiet-start prescription,
see section 2).

\begin{figure}
\epsscale{1} \plotone{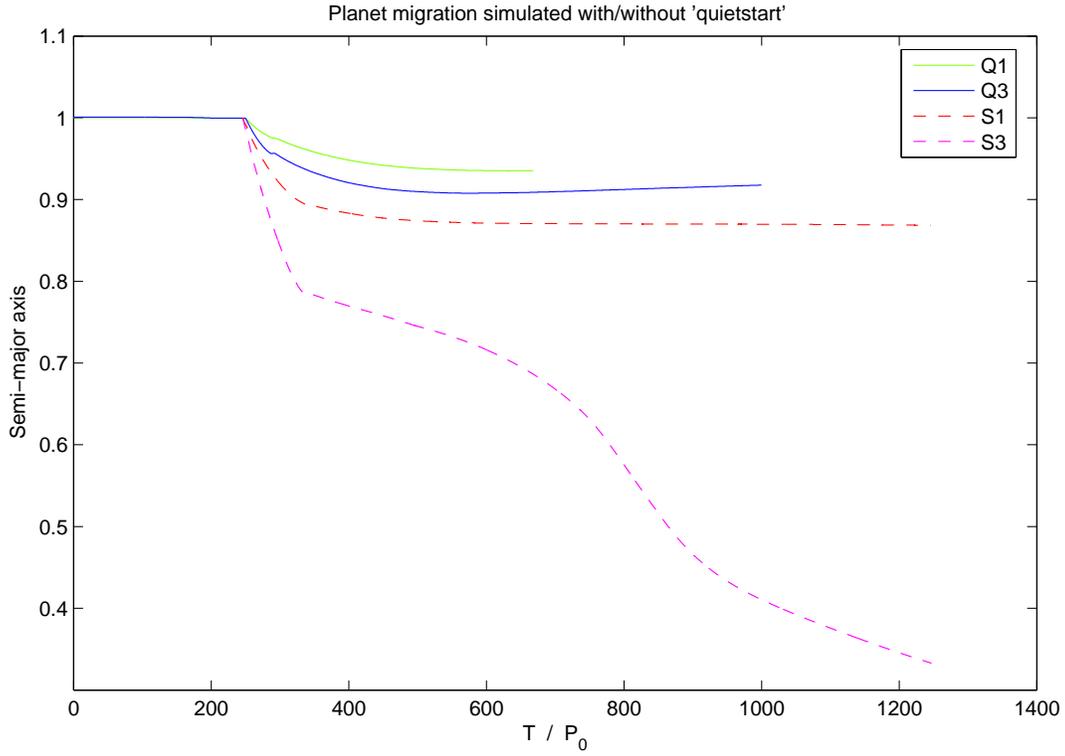} \caption{Migration with 'quiet start'.
From top to bottom, the curve according to Q1,Q3,S1,S3. First
$250P_0$ is 'quiet start' stage when the planet's orbit is fixed and
planet grows gently. \label{figure 9}}
\end{figure}

We adopt a quiet start in 8 models (See \textbf{Table}~\ref{table 1}),
four of them include the effect of self-gravitating while the others
do not. With a quiet start, a clear gap is formed near the planet's
orbit such that the corotation resonance is saturated from the onset.
Since the disk is able to establish a dynamical equilibrium through
adiabatic adjustments, the impulsive perturbation at the epoch of
planet release is minimized. In all cases, the migration rate is
greatly suppressed and type III migration is halted.  This result is
consistent with the MP conjecture that unsaturated corotation
resonances are responsible for inducing the type III migration.  (See
\textbf{Fig}.\ref{figure 9}).

\begin{figure}
\epsscale{1} \plotone{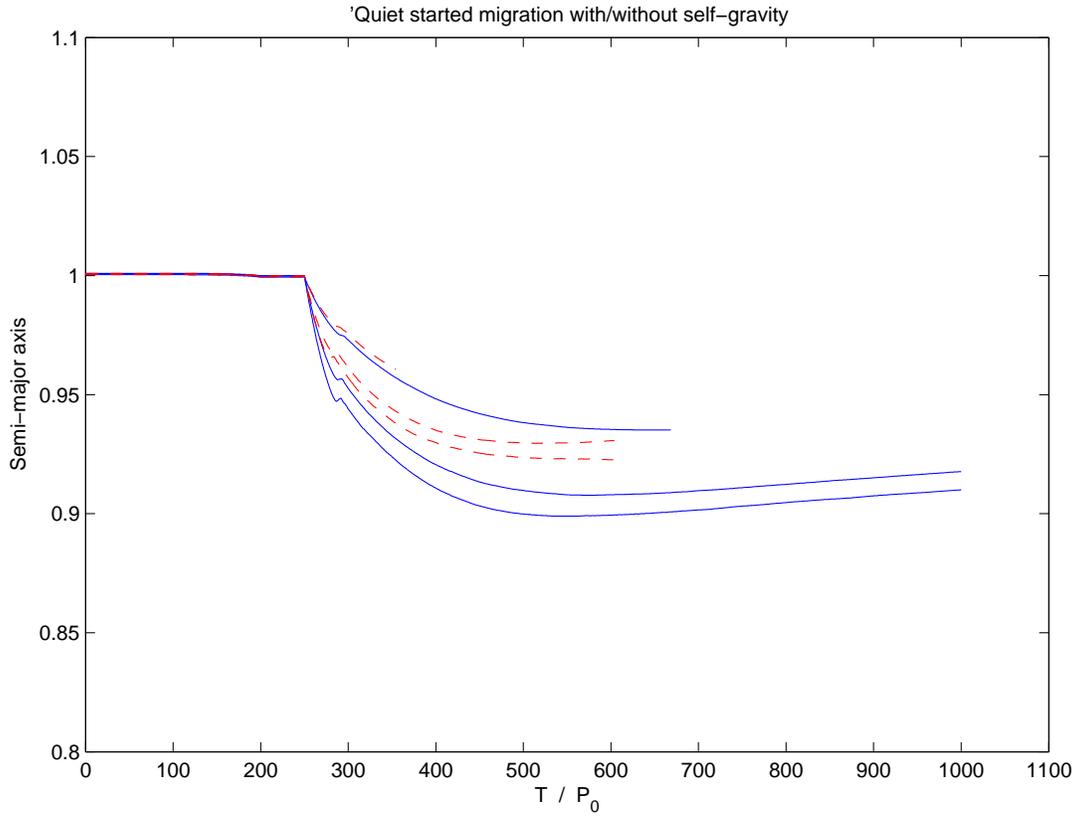} \caption{Migration in
self-gravitating disk. From top to bottom, the three red dashed
lines are QS1,QS3 and QS4 and the three blue solid lines are Q1,Q3
and Q4.  \label{figure 10}}
\end{figure}

The introduction of the quiet start algorithm reduces the
differential motion between the protoplanet and the disk gas at
this proximity and enhances the gravitational interaction between
them.  Consequently, the effect of self-gravity becomes more
pronounced earlier than it does in models without the quiet start
(See \textbf{Fig}.\ref{figure 10}).  With a quiet start and self
gravity of the disk, we minimize inconsistencies of the numerical
simulations for planets grow adiabatically in isolation.  In all
models with this combination, runaway migration is suppressed.

\subsection{Impulsive initial perturbation}
Although, run away migration is unlikely to be initiated
spontaneously, it can nevertheless be self sustained by mobile
planets.  Following the framework in MP's analysis, let us suppose
the planet has already acquired some initial velocity and is
moving inward relative to the disk gas.  The ``first move'' can be
the result of close-encounters between two gas giants or a strong
dynamical perturbation by some external stellar perturbation.

\begin{figure}
\epsscale{1} \plotone{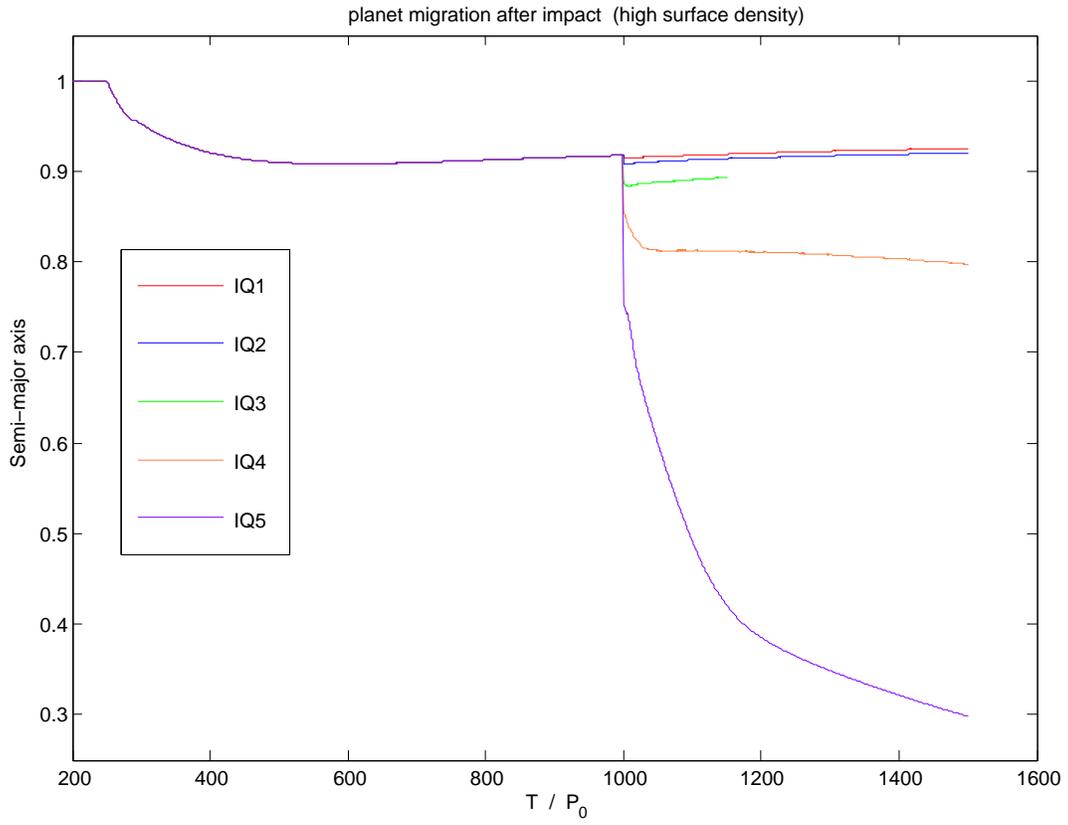} \caption{Migration after impact. It
shows the migration curves after close encounters(IQ1-IQ5). 'Runaway
migration' occurs when planet B is comparable to the A.
\label{figure 11}}
\end{figure}

In order to consider such a possibility, we simulated 5 additional
models which the disk and planetary parameters of model Q3. At the end
stage of Q3, a Jupiter-mass planet (A) is centered in a severely
depleted gap and its migration is essentially halted, --- {\it i.e.}
the planet-disk system has established an equilibrium structure.  At
this instant of time, we assume there is another planet (B) which
enters into a close encounter with planet A with an impact parameter
$0.5R_{roche}$ (where $R_{roche}$ is the Roche radius of planet A).
The duration of the close encounter is brief ($\sim P_{0}/20$).  In
the five test models (IQ1-IQ5), we adopt 5 different mass of planet B
(See \textbf{Table}.\ref{table 2}). The results are shown in
\textbf{Fig.}\ref{figure 11}.

The simulated results of models IQ1-IQ3 indicate that Planet A is
not significantly perturbed by close encounters with a much less
massive planet B. In each of these models, planet A move inward
slightly and then retain the dynamical equilibrium and its migration
is halted.  In model IQ4, planet B is sufficiently massive to induce
planet A to undergo a ten percent decay in its semi major axis (See
Fig.\ref{figure 11}). But planet A manages to open a sufficiently
clear gap that the corotation resonance become saturated.
Thereafter, runaway migration is also halted in model IQ4 but after
migration over a substantial radial extent.

\begin{figure}
\epsscale{1} \plotone{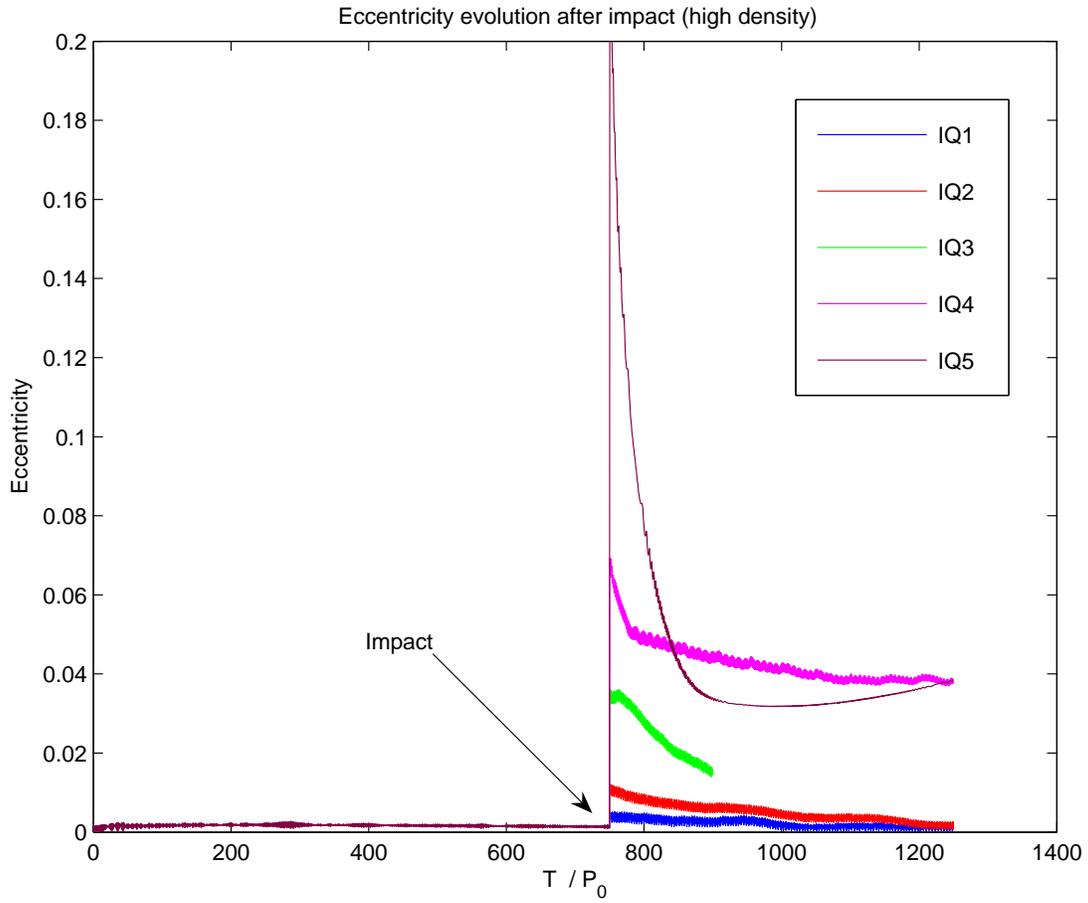} \caption{Eccentricity evolution after
impact. It shows the eccentricity evolution after close
encounters(IQ1-IQ5). The critical mass of the planet B which will
break the steady state of planet A is of Saturn mass and the
'runaway migration' occurs when planet B is comparable to the
A.\label{figure 12}}
\end{figure}

Shortly after the scattering event, planet A's eccentricity acquires
a finite amplitude (Fig.\ref{figure 12}).  In models IQ1-IQ3, planet
A acquires eccentricities smaller than the ratio of Roche radius to
its semi major axis.  Thus, planet A avoids direct contact with a
substantial amount of disk gas.  In model IQ4, planet A's modest
eccentricity shortly after the perturbation results in its periodic
excursion into the disk region beyond the gap.  In principle, the
disk gas is periodically fed to the planet and the condition for run
away migration is satisfied. (Although the corotation and Lindblad
resonances provide modest flux of angular momentum transport per
synodic period, these contributions accumulate in time.)  But the
tidal interaction between the disk gas and the planet through the
corotation resonance also leads to intense eccentricity damping
\citep{gol80,gol04} on time scale comparable to or shorter than that
for the runaway migration time scale (with the possible exception of
model IQ5 in which the planet's run away migration is launched).

Once the planet's eccentricity is suppressed, gas flow through the
co-orbital region, especially through the planet's Roche lobe, can
only be self sustained with a sufficiently large radial velocity.  Our
models show that finite amplitude perturbation excites a positive
feedback: 1) planet migrate inward leads to disk gas flowing pass it,
2) corotation resonance takes away the planet's angular momentum and
induces it to further migrate inwards.  The comparison between models
IQ3-IQ5 shows that the recoil speed of planet A increases with the
mass of planet B. The critical amplitude of the perturbation needed to
launch this self-sustained migration is that the supply into
co-orbital region, during the horseshoe orbits' libration time scale,
must be comparable to or larger than the mass of the planet (MP).

In model IQ5, the planet undergoes run-away migration over an
extensive radial distance.  In the co-moving frame of the planet, the
disk gas travels from inner to outer regions of the disk and remove
angular momentum from it through the corotation resonance (or
equivalently, the gas is being scattered to large distances by the
planet). The disk-planet interaction is greatly amplified during the
passage of the gas through planet's Roche lobe where the imbalance of
the torque is a direct consequence of a self-sustained potential
vorticity gradient. In comparison to model Q3, we find that the disk's
self-gravity, as a global and indirect effect, can only overwhelm the
local flow through the planet's Roche lobe when its vicinity is
severely depleted and the impact of the initial impulse has decayed or
is suppressed.

\begin{figure}
\epsscale{1} \plotone{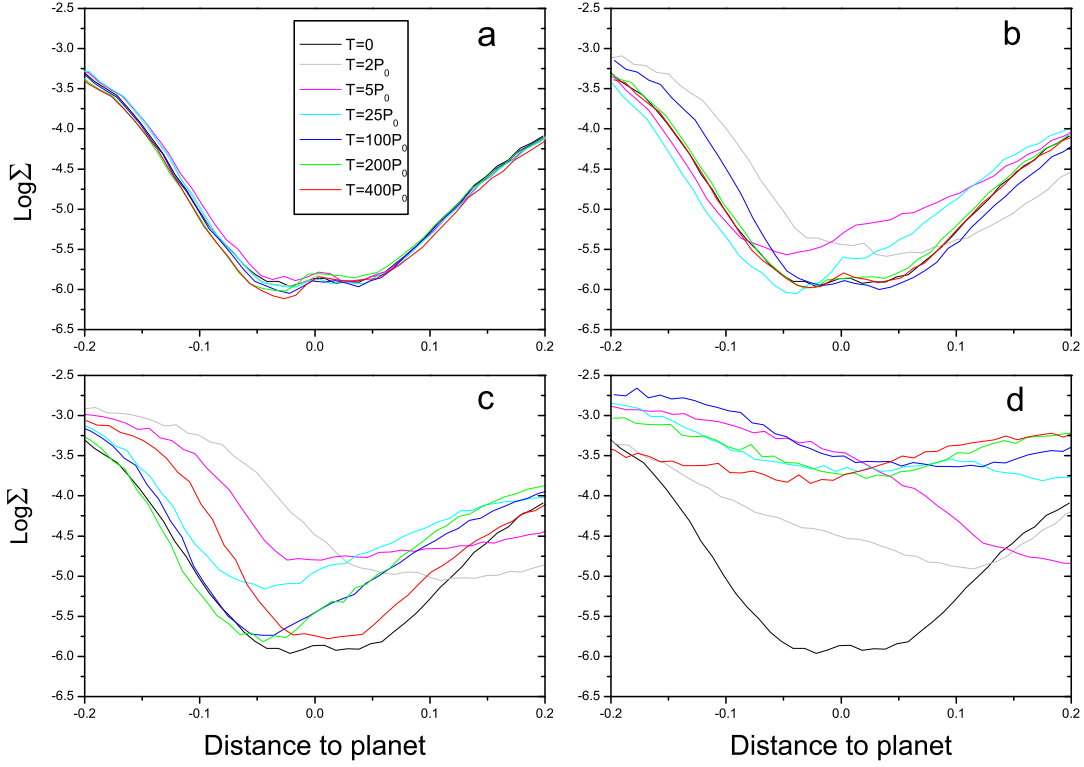} \caption{Logarithmic surface density
distribution in the vicinity of planet A after close encounter. From
the top left to the bottom right, figure a,b,c and d denote cases
named IQ1,IQ3,IQ4 and IQ5 respectively. Results are in co-moving
frame of planet A, X-axis is the distance to the planet A. The
region is $(-3R_{Roche},3R_{Roche})$,where $R_{Roche}=0.069$ is the
initial Roche radius of planet A in our unit. Colored curves denote
the different evolution time after close encounter: Black lines show
the equilibrium state before close encounter. \emph{a}. Planet A
remains its equilibrium since density profile within planet's
vicinity doesn't change much after small perturbation(IQ1 \& IQ2).
\emph{b}. Gas floods into planet A's vicinity short after relatively
large perturbation(The gray lines) but depletes soon after
$25P_0$(The light green lines). So there is no runaway
migration(IQ3). \emph{c}. Gas takes about $400P_0$ to deplete(The
red lines) in case IQ4 and planet A's fast migration is suppressed
finally since it can not self-sustained the vorticity gradient.
\emph{d}. Large radial perturbation allows planet A self-sustain the
vorticity gradient in its vicinity. So it keeps losing angular
momentum through the corotation resonance and under goes a run-away
migration. \label{figure 13}}
\end{figure}

We interpret the results in models IQ3-IQ5 in terms of torque due to
the corotation resonance from the gas inside the planet's radius. In
the discussion on models S3 and H2, we have already indicated that
inadequate resolution can introduce spurious torque imbalance which
may drive type III migration.  But we also showed in models SG3 and
Q3 that both self gravity and a quiet start can suppress type III
migration even when the flow is simulated with inadequate
resolution. In these cases, the torque imbalance may be further
reduced in simulations with more a refined numerical resolution.  In
comparison with the results of model IQ1-IQ3, the launch of type III
migration in models IQ4 and IQ5 is due to a physical effect rather
than a numerical flaw. In these models, the planet's orbital
evolution is dominated by the local torque (include those hasn't
been well resolved) when there is sufficient mass near it.
\textbf{Fig.}\ref{figure 13} shows the surface density distribution
within planet A's vicinity after the scattering event. The figures
a,b,c and d correspond to the cases IQ1,IQ3,IQ4 and IQ5
respectively. \emph{a}. For small perturbation, density profile
within the vicinity of planet doesn't change much and planet A
remains its equilibrium(IQ1 \& IQ2). \emph{b}. Gas starts to flood
into planet A's vicinity(The gray lines) short after relatively
large perturbation and evokes a potential vorticity gradient.
However the gas depletes soon after $25P_0$(The light green lines)
so there is no runaway migration and the planet come back to its
equilibrium shortly after perturbation(IQ3). \emph{c}. In IQ4 gas
takes about $400P_0$ to deplete(The red lines) and the corotation
resonance which associates with vorticity gradient was suppressed
then. Planet A had undergone fast migration through a extensive
radial region, however the migration is suppressed finally since it
can not self-sustained the vorticity gradient. \emph{d}. Large
radial perturbation allows planet A self-sustain the vorticity
gradient in its vicinity, and as a result of that, planet A's keep
losing angular momentum through the corotation resonance and under
goes a run-away migration.

\section{Summary and discussions}
In this paper, we are motivated to consider the origin and evolution
of the type-III (run away) migration.  This process is thought to be
important for Saturn mass planets which may have formed in disks more
massive than the minimum mass nebula.  We are particularly interested
in two issues: 1) whether this process can occur spontaneously for
isolated gas giants which formed through gas accretion onto solid
cores on time scales much longer than the dynamical time scale in the
disk; and 2) whether type-III migration can be self-sustained by a
planet which is strongly perturbed by a close encounter with another
planet.

We performed a series of numerical simulations to investigate the
orbit evolution of a embedded planet. In models S1-S4 which are of the
simplest settings in which a Jupiter-mass planet is inserted into an
isothermal disk with a constant surface density. In all cases,
migration occurs with a speed which is an increasing function of the
disk mass.  For disks with more than twice the mass of the minimum
mass nebula model, the planet undergoes type III migration.

Although these results are intriguing, we identify three technical
issues which may have led to an artificial outcome for the numerical
simulations.  In an attempt to carry out a convergence analysis, we
note that our low-resolution models (comparable in resolution to
most existing simulations) contain only $14\times14$ mesh grids
across the planet's Roche lobe. Insufficient numerical resolution
introduces an artificial torque imbalance especially for the gas
flow within the planet's Roche lobe.  When this region is intruded
by a disk flow, unphysical net torque is generated spuriously.  This
torque induces the planet to rapidly migrated.  When the flow
through the identical disks is simulated with twice the resolution
in each direction, the torque imbalance of the flows through the
Roche lobe and the rate of planetary migration are greatly reduced.
These studies suggest that under-resolved simulations may lead to
spurious run away migrations. In the work presented here, we cannot
yet demonstrate that we have reached numerical convergence. It will
require more powerful computational algorithms and tools to
determine the condition for adequate resolution.

The second technical problem which plagues many existing numerical
simulations is an inconsistent treatment of the disk self gravity.
In the computation of the force acting on the planet, both the
axisymmetric and the non-axisymmetric components of the gravity from
the disk gas are applied to the planet along with the host star's
gravity. But in the evaluation of the equation of motion of the disk
gas, the disk's own contribution to the gravity is not included.  In
low-mass disks, the discrepancy introduced by this approximation is
negligible.  But, in disks more massive than the minimum mass
nebula, this inconsistency can lead to a differential motion between
the planet and the disk gas near its orbit.  The absence of disk
self gravity also modifies the effect of angular momentum transport
across both the corotation and Lindblad resonances \citep{gol82}. A
comparison between models S3 and SG3 show that the extent of radial
migration after a time span of $10^3 P_0$ would be reduced by a
factor of two if the effect of the self gravity of the disk is
included. Nevertheless, runaway migration is not totally suppressed
by the disk's self gravity. Type III migration is spontaneously
launched in model SG4 which is also gravitationally unstable ---at
the out part of the disk $Q\sim1$. But in general, a self consistent
treatment of the disk self gravity can significantly slow down the
type-III migration rate.

The last technical issue we have considered is the artificial
initial conditions.  In the standard models S1-S4, the initial
motion of the disk gas is set up for Keplerian velocities so that it
is in a centrifugal balance with the host star's gravity.  At the
onset of the simulation, the introduction of the planet's gravity
induces a strong perturbation to the flow pattern.  Consequently,
the disk gas can easily enter into the Roche lobe of the planet and
intensely exchange angular momentum with the planet.  Since the
initial potential vorticity gradient is preserved, corotation
resonance is artificially intensified at the onset of the simulation
(see models SG1-SG4).  In the formation of the first-generation,
isolated gas giants, the planets' mass grows over many dynamical
time scales.  The dynamics and structure of the disk flow adjacent
to the planet adjust adiabatically through gap formation and
modification of stream lines. In an attempt to simulate this gradual
process, we carried out two series of models (Q1-Q4) with a
``quiet-start'' initial condition.  In these models, the mass of the
planet is increased gradually over $250 P_0$.  In the first series,
models Q1-Q4, we neglect the effect of the disk's self gravity
whereas in the second series, models QG1-QG4, the effect of the disk
self gravity is fully implemented.  In both sets of simulations, a
quiet start greatly suppress the corotation resonance and hence the
planet's type-III migration.  In all cases, (including the models
for a very massive disk Q4 and QG4), type III migration ceased.

While the quiet start initial condition is justified for the
first-generation, isolated planets, it is not appropriate for
strongly perturbed planets.  A large fraction of all known gas
giants reside in multiple planet systems. Indeed the formation of
first-generation planets promotes the build up of the cores and the
formation of gas giants planet beyond the outer edge of the gap
around them \citep{bry00}. Despite the gaseous background, if these
planets are formed with an initial separation less than about three
times the sum of their Roche radius, dynamical instabilities can
induce them to undergo close encounters well before the gas is
depleted \citep{zho07e}. In order to investigate these perturbation
in the presence of the disk gas, we simulated models IQ1-IQ5.  Our
results show that the runaway migration of a Jupiter mass planet can
be triggered by its close encounters with another planets more
massive than Saturn. Immediately after the encounter, the
eccentricity of the planet is excited such that it can undergo
radial excursion beyond the edge of the gap.  Although this motion
enables the disk gas to venture into the planet's Roche lobe, their
interaction through the corotation resonances damps the eccentricity
faster than directly induce type-III migration. However, the initial
impulse may be self sustained by the modest radial motion of the
planet.  In the co-moving frame of the planet, the disk gas moves
across its orbit, sustains a potential vorticity gradient, and
induces the planet to undergo type-III migration over large radial
distances.

In an attempt to account for the wide eccentricity distribution
among the extra solar planets, several authors have consider the
possibility of dynamical instability in multiple planet systems
\citep{pap01,jur07,cha07,zho07e}. The time scale for the onset of
dynamical instability is a rapidly increasing function of the
planet's separation. In many simulations, compact systems of planets
are imposed initially such that they become dynamically unstable on
time scale much shorter than both the growth time scale for gas
giant planets and the gas-depletion time scale (a few Myr) in their
nascent disks.  For example, Juric and Tremaine (2007) present
several models to show that, dynamical relaxation in a gas free
environment can induce the medium separation between gas giant
planets to increase from $<$5 to $>$12 within $10^{5-6}$yr. The
results presented here suggest that close encounters triggered by
dynamical instabilities, if frequently occurred prior to the
depletion of the disk gas, would launch proto gas giant planets on
type-III migration either towards their host stars or the outer edge
of their nascent disks. Either outcome may not be compatible with
the observed mass-period distribution of extra solar planets.  An
alternative scenario is the formation of first generation gas giants
strongly modified their neighborhood by opening up planetesimal gaps
at several Hill's radius from themselves \citep{zho07} as well as
wide gas gaps \citep{bry00}. With moderate large separations, the
growth time scale for dynamical instabilities may be lengthened by
one or more orders of magnitude. Provided the dynamical instability
leads to close encounters between gas giants after the depletion of
the disk gas, it is possible for most of them to remain in the
proximity of their birth place. Quantitative verification of this
conjecture will be presented elsewhere.

\section{Acknowledgement}
We thank Drs Hui Li and Jilin Zhou for useful conversation.  The
work is in parts supported by a grant from National Basic Research
Program of China (2007CB4800),Natural Science Foundation of China
(10403004),National Science Council,Taiwan NSC94-2752-M-001-002-PAE,
NASA (NAGS5-11779, NNG04G-191G, NNG06-GH45G), JPL (1270927), and
NSF(AST-0507424).

{}

\clearpage

\begin{deluxetable}{cccccc}
\tablecaption{Simulations in this paper\label{table 1}}
\tablehead{\colhead{Case} &\colhead{Disk Surface Density} &
\colhead{Resolution} & \colhead{Self-gravity}&\colhead{Quiet
start}& \colhead{Runaway migration}} \startdata
S1   & $0.6\times10^{-3}$ &  $512\times512$  & No & No & No\\
S2   & $0.9\times10^{-3}$ &  $512\times512$  & No & No & No\\
S3   & $1.2\times10^{-3}$ &  $512\times512$  & No & No & Yes\\
S4   & $1.5\times10^{-3}$ &  $512\times512$  & No & No & Yes\\
\tableline
H1   & $0.6\times10^{-3}$ & $1024\times1024$ & No & No & No\\
H2   & $1.2\times10^{-3}$ & $1024\times1024$ & No & No & No\\
\tableline
SG1  & $0.6\times10^{-3}$ &  $512\times512$  & Yes & No & No\\
SG2  & $0.9\times10^{-3}$ &  $512\times512$  & Yes & No & No\\
SG3  & $1.2\times10^{-3}$ &  $512\times512$  & Yes & No & No\\
SG4  & $1.5\times10^{-3}$ &  $512\times512$  & Yes & No & No\\
\tableline
Q1   & $0.6\times10^{-3}$ &  $512\times512$  & No & Yes & No  \\
Q2   & $0.9\times10^{-3}$ &  $512\times512$  & No & Yes & No  \\
Q3   & $1.2\times10^{-3}$ &  $512\times512$  & No & Yes & No  \\
Q4   & $1.5\times10^{-3}$ &  $512\times512$  & No & Yes & No  \\
\tableline
QG1  & $0.6\times10^{-3}$ &  $512\times512$ & Yes & Yes & No  \\
QG2  & $0.9\times10^{-3}$ &  $512\times512$ & Yes & Yes & No  \\
QG3  & $1.2\times10^{-3}$ &  $512\times512$ & Yes & Yes & No  \\
QG4  & $1.5\times10^{-3}$ &  $512\times512$ & Yes & Yes & No  \\

\enddata

\end{deluxetable}


\begin{deluxetable}{cccccc}
\tablecaption{Simulations of Impact \label{table 2}}
\tablehead{\colhead{Case} &\colhead{Encounter planet mass}&
\colhead{Runaway migration}} \startdata
IQ1   & $5 M_{earth}$ & No  \\
IQ2   & $15M_{earth}$ & No  \\
IQ3   & $50M_{earth}$ & No  \\
IQ4   & $100M_{earth}$& Critical  \\
IQ5   & $300M_{earth}$& Yes  \\

\enddata

\end{deluxetable}

\clearpage

\clearpage

\begin{thebibliography}{}
\bibitem[Balmforth \& Korycansky 2001]{bal01} Balmforth, N. J.,Korycansky,D.G. 2001,\mnras,326,833
\bibitem[Bryden {\it et al.} 2000]{bry00} Bryden,G.,R¨®?yczka,M.,Lin,D.N.C.,\& Bodenheimer,P. 2000,\apj,540,1091
\bibitem[Chatterjee,Ford,\& Rasio 2007]{cha07} Chatterjee,S.,Ford,E.B.,\& Rasio,F.A. 2007,\apj,submitted
\bibitem[Cresswell \& Nelson 2006]{cre06} Cresswell,P., \& Nelson,R.P. 2006,\aa,450,833
\bibitem[Crida \& Morbidelli 2007]{cri07} Crida,A.,\& Morbidelli,A. 2007, \mnras,377,1324
\bibitem[D'Angelo,Bate \& Lubow 2005]{ang05} D'Angelo,G., Bate,M.R., \& Lubow,S.H. 2005, \mnras,358,316
\bibitem[D'Angelo {\it et al.} 2006]{ang06} D'Angelo,G., Lubow,S.H., \& Bate,M.R. 2006, \apj,652,1698
\bibitem[de Val-Borro {\it et al.} 2007]{dev07} de Val-Borro,M., Artymowicz,P., D'Angelo,G.,\& Peplinski,A.,2007,\aa,in press
\bibitem[Gammie 1996]{gam96} Gammie, C.F. 1996,\apj,462,725
\bibitem[Godon 1996]{god96} Godon, P. 1996,\mnras,282,1107
\bibitem[Goldreich \& Tremaine 1979]{gol79} Goldreich,P., \& Tremaine,S. 1979,\apj,233,857
\bibitem[Goldreich \& Tremaine 1980]{gol80} Goldreich,P., \& Tremaine,S. 1980,\apj,241,425
\bibitem[Goldreich \& Tremaine 1982]{gol82} Goldreich,P., \& Tremaine,S. 1982,ARA\&A,20,249
\bibitem[Goldreich {\it et al.} 2004]{gol04} Goldreich, P., Lithwick, Y.,\& Sari,Re'em, 2004,\apj,614,497
\bibitem[Ida \& Lin 2004]{ida04a} Ida, S.,\& Lin, D.N.C. 2004,\apj,604,388
\bibitem[Ida \& Lin 2004]{ida04b} Ida, S.,\& Lin, D.N.C. 2004,\apj,616,567
\bibitem[Ida \& Lin 2005]{ida05} Ida, S.,\& Lin, D.N.C. 2005,\apj,626,1054
\bibitem[Ida \& Lin 2007]{ida07} Ida, S.,\& Lin, D.N.C. 2007,\apj,submitted
\bibitem[Ida {\it et al.} 2000]{ida00} Ida, S., Bryden, G., Lin, D. N. C.,\& Tanaka,H. 2000,\apj,534,428
\bibitem[Ivanov,Papaloizou \& Polnarev 1999]{iva99} Ivanov,P.B.,Papaloizou,J.C.B, \& Polnarev,A.G. 1999,\mnras,307,79
\bibitem[Juric \& Tremaine 2007]{jur07} Juric, M.,\& Tremaine,S. 2007,\apj,submitted
\bibitem[Koller 2003]{kol03} Koller, J., Li, H.,\& Lin,D.N.C. 2003,\apj,596,91
\bibitem[Koller 2004]{kol04} Koller, J. 2004,Thesis (PhD). RICE UNIVERSITY, Source DAI-B 65/02, p.789
\bibitem[Laughlin,Steinacker \& Adams 2004]{lau04} Laughlin, G., Steinacker, A.,\& Adams,F. C. 2004,\apj,608,489
\bibitem[Li {\it et al.} 2005]{li05} Li,H.,Li,S.T.,Koller,J.,Wendroff,B.B.,Liska,R.,Orban,C.M., Liang,E.P.T.,\& Lin, D.N.C. 2005,\apj,624,1003
\bibitem[Lin \& Papaloizou 1986]{lin86} Lin,D.N.C. \& Papaloizou,J.C.B. 1986,\apj,309,846
\bibitem[Lin \& Papaloizou 1993]{lin93} Lin,D.N.C. \& Papaloizou,J.C.B. 1993,in Protostars and planets III,ed.E.H.Levy \& J.I.Lunine (Tucson:Univer.Arizona Press),749
\bibitem[Lin,Bodenheimer \& Richardson 1996]{lin96} Lin,D.N.C.,Bodenheimer,P.,\& Richardson, D.C. 1996,Nature,380,606
\bibitem[Masset \& Papaloizou 2003]{mas03} Masset,F.S. \& Papaloizou,J.C.B. 2003,\apj,588,494
\bibitem[Masset {\it et al.} 2006a]{mas06a} Masset, F. S., Morbidelli, A., Crida, A.\& Ferreira, J. 2006,\apj,642,478
\bibitem[Masset,D'Angelo \& Kley 2006b]{mas06b} Masset, F. S., D'Angelo, G.\& Kley,W. 2006,\apj,652,730
\bibitem[Nelson \& Benz 2003]{nel03a} Nelson, A. F.\& Benz,W. 2003, \apj, 589,556
\bibitem[Nelson \& Benz 2003]{nel03b} Nelson, A. F.\& Benz,W. 2003, \apj, 589,578
\bibitem[Nelson \& Papaloizou 2004]{nel04} Nelson, R.,\& Papaloizou, J.C.B. 2004,\mnras,350,849
\bibitem[Ogilvie \& Lubow 2006]{ogi06} Ogilvie,G.I. \& Lubow,S.H.2006,\mnras,370,784
\bibitem[Papaloizou \& Lin 1989]{pap89} Papaloizou, J. C. B.,\& Lin, D. N. C. 1989,\apj,344,645
\bibitem[Papaloizou \& Terquem 2001]{pap01} Papaloizou, J. C. B \& Terquem,C. 2001,\mnras,325,221
\bibitem[Papaloizou \& Terquem 2006]{pap06} Papaloizou, J. C. B \& Terquem,C. 2006,RPPh,69,119
\bibitem[Pollack {\it et al.} 1996]{pol96} Pollack, J. B., Hubickyj, O., Bodenheimer, P., Lissauer,J. J., Podolak, M.,\& Greenzweig, Y. 1996, \icarus, 124,62
\bibitem[Tanaka,Takeuchi \& Ward 2002]{tan02} Tanaka, H., Takeuchi, T.,\& Ward,W. R. 2002,\apj,565,1257
\bibitem[Thommes \& Murray 2006]{tho06} Thommes, E. W.,\& Murray, N. 2006,\apj,644,1214
\bibitem[Ward 1984]{war84} Ward,W.R. 1984,in Planetary rings,ed. R.Greenberg and A.Brahic,(Tucson:University of Arizona Press),pp.660-684
\bibitem[Ward 1991]{war91} Ward,W.R. 1991,Lunar \& Planet.Sci.,22,1463
\bibitem[Ward 1992]{war92} Ward,W.R. 1992,Lunar \& Planet.Sci.,22,1491
\bibitem[Ward 1997]{war97} Ward,W.R. 1997,\icarus,126,261
\bibitem[Yuan \& Yen 2005]{yua05} Yuan,C., \& David,C.C. Yen 2005,JKAS,38,197
\bibitem[Zhou \& Lin 2007]{zho07} Zhou, J.L., \& Lin, D.N.C. 2007,\apj,in press
\bibitem[Zhou {\it et al.} 2005]{zho05} Zhou, J.L., Aarseth,S.J., Lin,D.N.C., \& Nagasawa,M. 2005,\apj,631,85
\bibitem[Zhou,Lin \& Sun 2007]{zho07e} Zhou, J.L., Lin, D.N.C., \& Sun,Y.S. 2007,\apj,in press
\end{thebibliography}
\end{document}